# Current-modulation annealing to control microwave permittivity in composites with melt-extracted microwires

Y. Luo,[1] H. X. Peng,[2, (a)] F. X. Qin,[2, (b)], J.S. Liu,[3] H. Wang,[2] F. Scarpa,[1] B. J. P. Adohi,[4] and C. Brosseau[4]

[1]*Advanced Composite Centre for Innovation and Science, Department of Aerospace Engineering, University of Bristol, University Walk, Bristol, BS8 1TR, UK*
[2]*Institute for Composites Science and Innovation (InCSI), School of Materials Science and Engineering, Zhejiang University, Hangzhou, 310027, China*
[3]*School of Materials Science and Engineering, Inner Mongolia University of Technology, China*
[4]*Lab-STICC, Université de Brest, 6 avenue Le Gorgeu, CS 93837, 29238 Brest Cedex 3, France*

To whom the correspondence should be addressed.
(a) hxpengwork@zju.edu.cn
(b) faxiangqin@zju.edu.cn

**Abstract**

We investigate the microwave properties of epoxy-based composite containing melt-extracted CoFeBSiNb microwires fabricated by a combined current-modulation annealing (CCMA) technique. We observe a shift of the resonance peak in the effective permittivity spectra of the composite sample containing annealed 25 mm Nb-doped microwires as an applied magnetic field is increased. This observation is consistent with the absorption-dominated impedance for thick microwires and the ferromagnetic resonance phenomenon. It is shown that CCMA is an appropriate technique to release internal residual stresses. Hence, for samples containing small amounts of Nb, we observe that CCMA allows us to suppress the high frequency resonance peak observed in samples containing as-cast wires. However, for samples containing a high amount of Nb, the high frequency peak remains despite the CCMA treatment. In this case, the observation of a two-peak feature in the permittivity spectra is attributed to the coexistence of the amorphous phase and a small amount of nanocrystallites distributed at the wire surface. However, due to large magnetostatic energy of long (35 mm) and short (15 mm) as-cast wires and imperfect wire-epoxy bonding no shift of the resonance peak and the characteristic double peak of the permittivity spectrum can be detected. Overall, CCMA emerges as a promising strategy to control microwave permittivity in composites with melt-extracted microwires.

# I. INTRODUCTION

Amorphous ferromagnetic microwires are interesting materials both for fundamental research and for technological applications. They have been extensively studied to provide materials for magnetic and stress sensors.[1,2] Among the existing microwire families, melt-extracted (MET) microwires possess superior mechanical and soft magnetic properties due to their ultra-high cooling rate in the fabrication stage compared to glass coated microwires.[3,4] The search for more efficient tunable sensing and frequency-agile materials has led to the investigation of polymer-based composites containing MET wires.[5,6] Specific attention has been paid to their collective response to external magnetic fields and mechanical tension arising from the giant magnetoimpedance (GMI) /giant stress-impedance (GSI) effects.[1,2] Moreover, it was demonstrated that geometrical factors, such as wire alignment[7] and aspect ratio which controls the demagnetization factors,[8] can significantly influence their overall dielectric behavior. It was shown that the wire length can impact the microwave behavior of Fe-based wire composites which manifests crossover field phenomenon and linear field-tunability.[9]

In another perspective, one should note that the intrinsic electromagnetic (EM) properties of microwires relating to their domain structure and microstructure control the effective microwave properties of the composites. Generally, there are two ways to optimize the EM performance of individual microwires. Firstly, it is known that the doping of CoFe-based microwires by metallic elements such as Cu, Cr, Zr, Nb induces excellent soft magnetic behavior.[10,11]. Nb has drawn significant interest in this regard, as early experiments revealed excellent mechanical and soft magnetic properties.[12,13,14] Secondly, several post-annealing techniques have together provided a great deal of information about the control and tunability properties on the microwire properties in response to incident EM waves.[15] For example, current annealing has been extensively studied using DC,[16,17] AC,[18] and pulse current (PC).[19] It has been ascertained that the annealing process initiates internal stress relaxation improving circumferential permeability which produces enhancement to the field/stress sensitivity.[17] However, conventional current annealing techniques have their own limitations: DC annealing can stabilize the circumferential anisotropy but is likely to generate excessive heat that can possibly induce damage to the wires; PC annealing fails to provide the persistent power that is required to improve the domain structure. To overcome these difficulties, a combination of DC and PC annealing techniques, named the combined current-modulation annealing (CCMA), has been shown to provide a good compromise.[20] CCMA can increase the thermal activation energy during the PC annealing stage while requiring moderate energies during the DC annealing stage for the formation of uniform domain structures, leading to sensitive and flexible GMI response. Understanding the influence of CCMA on the microwave properties of polymer composites containing microwires is quite challenging since the wire-polymer interfacial stress mechanism still remains intricate not only from its physical origin but also from its lack of control; evidence of the interdependence between wire length, annealing, and EM properties of such composites have been rarely revealed.[21] One major challenge is how to avoid dielectric losses-especially in large-volume structures.

With this background, we wish to examine the effective microwave properties of polymer composites containing MET ferromagnetic Co-based microwires subjected to external magnetic field. The interdependence between CCMA, chemical composition, microwire length and microwave response of polymer composite samples is demonstrated. In this work, four very interesting effects emerge. (i) We observe a shift of the resonance peak in the effective permittivity spectra of the composite sample containing annealed 25 mm Nb-doped microwires as the magnetic field is increased. This observation is consistent with the absorption-dominated impedance for thick microwires and the ferromagnetic resonance. (ii) The permittivity spectrum associated with composites containing as-cast wires exhibits a peak close to 5 GHz. We present evidence that this peak is due to the quasi-longitudinal anisotropy field and heterogeneously distributed stress induced during the fabrication stage of the wire. (iii) A two-peak feature has been identified in the effective permittivity spectrum of the sample containing 25 mm microwires with a high amount of Nb. The high frequency peak remains after CCMA and high resolution transmission electron microscopy (HRTEM) observations suggest that it is related to surface modification, *i.e.* nanocrystallites embedded in the continuous amorphous phase of wires. (iv) CCMA can significantly improve the permittivity results for composite samples containing as-cast long (35 mm) and short (15 mm) wires even if their high magnetostatic energy and poor interfacial bonding have detrimental effects.

## II. SYNTHESIS AND CHARACTERIZATION OF SAMPLES

Samples of amorphous $Co_{69.25}Fe_{4.25}B_{13.5-x}Si_{13}Nb_x$ microwires (nominal values of $x$=0,1,3) with average diameter of 45 $\mu$m were synthesized via the MET technique. Details of the fabrication for producing the microwires may be found in Ref. [3]. Fig. 1 presents the SEM morphology of as-cast $Co_{69.25}Fe_{4.25}B_{12.5}Si_{13}Nb_1$ wires. No cracks or defects are observed along the wire surface and the cross sectional area also indicates smooth perimeter. The tensile strength of Nb-containing Co-based MET wires can be as high as 4k MPa accompanied by a vein pattern from the fracture imaging (inset of Fig. 1), suggesting excellent mechanical properties.[14]

CCMA was performed by applying a PC at 50 Hz with amplitude of 90 mA during 480s, followed by a DC with amplitude 65 mA passing through the wires for 480 s (inset of Fig. 1). It is worth noting that the DC annealing time and magnitude should be carefully controlled to avoid partial burning of wires. This is done by experimental measurement and numerical calculation of the transient temperature in the temperature ramp-up stage associated with their thermal capacity.[20] Then, as-cast and annealed wires with lengths of 15, 25 and 35 mm were randomly dispersed into epoxy (PRIME$^{TM}$ 20LV, Gurit UK), followed by a standard curing cycle. For the microwave characterization, the sample have dimensions of $70\times10\times1.8$ mm$^3$.[5] For comparison, a specimen containing $Co_{69.25}Fe_{4.2}B_{13.5}Si_{13}$ microwires was also fabricated by the same method. Composite samples containing microwires are noted as Nb0 ($x$=0), Nb1 ($x$=1), and Nb3 ($x$=3), respectively.

The samples were structurally characterized by HRTEM (JEM 2010F). The HRTEM samples were prepared using a Gatan 691 ion beam thinner. The effective complex (relative) permittivity $\varepsilon = \varepsilon' - j\varepsilon''$ was obtained from the measurement of the S parameters in the frequency range from 0.3 to 6 GHz using a vectorial network analyzer (Agilent, model H8753ES) and SOLT (short, open, load, through) calibration of an asymmetric microstrip transmission line containing the sample.[22] Software is employed to convert the S parameters into the complex permittivity of the material. The electromagnetic measurement was carried out with a wave vector of the electromagnetic field perpendicular to the wires. The quasi-TEM transverse electromagnetic mode, which is the only mode that propagates in the structure, makes the analysis of the complex transmission and reflection coefficients relatively simple. It is worth stressing that all sample thicknesses and internal characteristic lengths of the heterogeneities are much smaller than the wavelength of the electromagnetic wave probing the material samples. This suggests that, within the quasi-static approach, the dominant loss mechanism comes from the absorption, rather than from the inhomogeneities scattering. An error analysis indicates systematic uncertainties in $\varepsilon'$ (<5%) and $\varepsilon''$ (<1%) for the data.[22] An external magnetic bias is swept from 0 to $\pm 5$ kOe. Magnetic field measurements of $\varepsilon$ were performed by placing the transmission line between the poles of an electromagnet to provide the dc magnetic field, *H*, which was monitored by using a Gauss meter equipped with a Hall element. All measurements reported here were performed at room temperature.

### III. RESULTS AND DISCUSSION

#### A. Influence of CCMA

Figure 2 displays the effective complex permittivity spectra of sample Nb1 containing 25 mm as-cast and CCMA microwires. A low frequency peak is identified around 4 GHz in the composite containing as-cast microwires and zero magnetic field (Fig. 2(c)), which is also seen in the $\varepsilon''$ spectra of samples annealed with CCMA (Fig. 2(d)). When a magnetic bias is applied, a high frequency peak in the spectra of composite samples containing as-cast wires is observed in the range of 5.0-5.5 GHz depending on the nominal field value. However, such a peak is suppressed in the specimen containing CCMA wires (Fig. 2(d)) except for the highest values of the applied field. The experimental data indicate no striking correlation between the peak positions and the applied magnetic field (Figs. 2(b) and 2(d)).

The observed low frequency peak is related to dipolar resonance whose spectral position can be determined using $f = c/2l\sqrt{\varepsilon_m}$, where $c$, $l$ and $\varepsilon_m$ are respectively the speed of light in vacuum, wire length, and the permittivity of the matrix.[21] Taking $l$=25 mm and $\varepsilon_m$ =2.5,[5] $f$ is calculated to be 3.8 GHz, which is close to the observed resonance peak. The residual discrepancy is attributed to the surface defects during the fabrication stage.[3] Now to address the physical origin of the high frequency peak, it is first important to observe that the present as-cast MET Co-based microwires have magnetostatic properties which resemble those of Fe-based wires. Shen *et al.* reported that the ratio of the remanence to saturation magnetization of MET Co-based wires is $\approx 0.42$,[18] which is close to the value found for glass coated Fe-based.[23] This indicates a quasi-longitudinal anisotropy of the MET wires, suggesting that the high frequency peak is related to the natural ferromagnetic resonance (FMR), *i.e.* $f_{FMR} = \gamma(M_s + H_a/2\pi)$, where $\gamma$, $M_s$, and $H_a$ ($H_a = M_s$ for Fe-based wires) are the gyromagnetic ratio, saturation magnetization and anisotropy field, respectively.[24] As the external field is increased, Kittel's relation predicts that this peak position should shift towards higher frequencies.[25] However, there is no clear experimental evidence for such a blue shift as per Figs. 2(c) and 2(d). To account for this discrepancy, one needs to consider the wire fabrication process. During the wire fabrication, a microwire is directly melt-extracted from the instantaneous contact between the copper wheel and molten puddle. Wang and co-workers reported that the temperature distribution varies drastically from the wheel tip to the puddle bottom, leaving the wire surface highly amorphous due to the maximum cooling rate.[3] This is also placed in evidence by the HRTEM image of the as-cast Nb1 wires where a typical amorphous morphology can be shown in addition to a halo from the FFT pattern (Fig. 3(a)).

During wire fabrication, the massive temperature drop near the wheel tip is likely to cause significant microstructural changes at the wire surface and consequently residual stresses increase which manifest as an additional contribution to the anisotropy field $H_a$.[26,27]

At this point it might be appropriate to understand the role of CCMA since the microstructure of wires treated by CCMA is still amorphous (Fig. 3(b)). The relaxation of residual stresses allows a consistently thermal modification of domain structure and hence a stabilized longitudinal anisotropy field $H_a$. Note that the modified longitudinal anisotropy field is less than the quasi-longitudinal field of as-cast wires. This explains why the FMR peak merges with the dipole resonant peak at low frequencies between 3 and 4 GHz after CCMA treatment (Fig. 2(d)). In this context, it is also worth noting that the stress inhomogeneities in the Co-based microwires degrades the outer shell of the bamboo-like domain structure, inducing a change of sign of the anisotropy constant.[28] The overlapped frequency range of the FMR and dipole resonance peaks (close to 4 GHz) of annealed wires will be discussed later.

We now substantiate the influence of CCMA by providing a simple, intuitive, understanding of its effect. As shown in Figs. 2(b) and 2(d), the magnitude of both real and imaginary parts of the effective permittivity at a given frequency for composites containing annealed wires is smaller than the corresponding values for the composites with as-cast wires. This is attributed to the stresses exerted from the epoxy matrix. For the composite containing as-cast wires, a compressive stress is exerted onto the inhomogeneous internal stress concentration sites at wire surface causing a tensile stress along the longitudinal direction.[30,31] In an earlier study,[32] it was shown that this longitudinal stress increases the magnetic permeability of Co-based wires which enhances the stress-induced impedance and hence the effective permittivity. In sharp contrast, because CCMA releases residual stress, less mechanical load is transferred to microwires, resulting in reduced stress-induced impedance and effective permittivity.

The strong variation in the overall scattering parameters $S_{11}$ and $S_{21}$ displayed in Fig. 4 is reminiscent to the results presented in a previous study,[9] which explains that such phenomenon is a consequence of the competition between GMI and FMR. However, the blueshift of the low frequency peak at low field (Figs. 4(b) and 4(d)) contrasts with the observed redshift feature observed in Fe-based wire composites.[9] However, we should note that the dielectric response of the samples containing the present microwires having a much larger diameter of 45 $\mu$m, as opposed to that of samples with Fe-based glass coated wires (16.8 μm of the metallic core) in Ref [9], is dominated by its absorption coefficient. Recall that the $S_{21}$ coefficient is determined by the real part of the impedance (Re($Z$)), whereas the $S_{11}$ and absorption coefficient are related to its imaginary part (Im($Z$)). Calculations,[33,34] performed for an array of thick ferromagnetic wires showed that Re($Z$) decreases and Im($Z$) increases as frequency in increased. Consistent with these calculations, the low field absorption is enhanced and determines the blue shift of the low frequency peak shown in Figs. 4(b) and 4(d). If the magnetic field is increased above 300 Oe, the reactance-induced absorption is saturated and the FMR (occurring between 3 and 4 GHz) which has a predominant role, shifts to higher frequencies according to Kittel's relation.[25] In summary, the field-dependent blueshift resonance peak positioned in the frequency range 3-4 GHz is a consequence of the competition between the low field absorption and the FMR at high magnetic field.

### B.     Influence of Nb doping

In Figures 5 and 6 we show the effective permittivity spectra of Nb0 and Nb3 composite samples, respectively. Figure 5 displays the complex permittivity of Nb0 sample containing 25 mm CCMA wires in the frequency range from 3 to 6 GHz. The results are very similar to the dielectric characterization of Nb1 sample analyzed in Sect. A. Fig. 5 indicates that the high frequency peak in the permittivity spectrum was again suppressed by using CCMA (*For brevity, permittivity spectrum for as-cast Nb0 specimen is not shown here*). In Fig. 6 we show the effective permittivity spectra of the Nb3 sample containing either as-cast or CCMA 25 mm microwires. The salient feature of these experimental results is that the two-peak structure of the $\varepsilon''$ spectra is evidenced for the Nb3 samples containing both as-cast (Fig. 6(c)) and CCMA (Fig. 6(d)) wires.

We argue that the key to understand the unique high frequency permittivity signature of the composite containing Nb3 wires is related to the microstructure change of the microwires with Nb doping. It has been already demonstrated that the MET CoFeSiB wires possess a core-shell (CS) structure with an amorphous core and a thin nanocrystalline shell.[6] Two comments are in order. Firstly, the inhomogeneously localized residual stresses initiate a nanocrystallite nucleation process. Secondly, the rapid cooling rate employed in the fabrication procedure inhibits the growth of nano-grains. In addition, it is worth observing that this nanocrystalline phase is strongly metastable. It is well known that Nb is used to enhance the glass forming ability and mechanical properties of metallic glasses.[35] Moreover, Nb is an efficient inhibitor for crystalline growth as it is rejected from the crystallization front to the amorphous phase due to its smaller diffusivity arising from the relatively larger atom size. It is likely that wires doped with 1 % of Nb or CoFeSiB wires are unable to generate nanocrystallites due to an insufficient number of nucleation sites arising from stress concentration locations. However, for doping content higher than 3%, we rely on past studies which demonstrated that Nb atoms are well known to maintain the thermal stability of nanocrystallites localized at nucleation sites with high internal-stress at wire surface.[36,37] The effective permittivity of such CS structure can be described by $\varepsilon=\beta\varepsilon_{amor}+(1-\beta)\varepsilon_{nano}$, where $\beta$ is the relative weight ratio of the amorphous phase, $\varepsilon_{amor}$ and $\varepsilon_{nano}$ indicating the intrinsic permittivity of the amorphous and nanocrystalline phases of microwires, respectively.[6] This hypothesis is validated by the HRTEM image of as-cast Nb3 microwires (Fig. 7(a)), which displays a small nanocrystalline phase of ~2 nm in size embedded in the amorphous region. As also observed in Fig. 7(b) corresponding to a CCMA Nb3 sample, a larger amount of this nanocrystalline phase is formed compared with the as-cast sample and is uniformly embedded in the amorphous phase together with a typical polycrystalline ring detected in the FFT pattern. The average size of the nanocrystalline phase is close to 1.5 to 2 nm. This observation further confirms the role of Nb as enhancer of the crystalline thermal stability. As magnetic field is increased the high frequency resonance peak blueshifts in the frequency range of 5-5.5 GHz, which is consistent with results described in Sect. A. We expect that a more important nanocrystalline phase on wires would, as a side effect, hinder the overall dielectric response of the composite sample since nanocrystallization generally has the effect of diminishing the soft magnetic properties to some extent.

In order to qualitatively capture the experimental results, notably the observed two-peak feature, it should be mentioned that nanocrystallites are likely to aggregate as discontinuous nano-clusters (nano-regions observed in Fig. 7(b)) since Nb atoms tend to accumulate at the wire surface. Hence, the dielectric contribution of these nanocrystallites can be regarded as arising from an array of randomly dispersed short "thin needles". The Nb3 composites can be also considered as a non-conductive medium filled with two randomly scattered arrays, *i.e.* amorphous microwires and nanocrystallite needles (as tentatively depicted in Fig. 8). We suggest that an effective length $l_{\text{eff}}$ accounts for the equivalent dielectric response of these two kinds of objects. Hence, the dipole resonance frequency should read $f = c / 2\, l_{\text{eff}} \sqrt{\varepsilon_{\text{m}}}$, where the $l_{\text{eff}}$ of nanocrystallites is much smaller than that of the amorphous phase, thus rendering possible the high frequency resonance. In summary, the average response of the amorphous and nanocrystalline phase of the microwires potentially explains the low and high frequency peaks in the permittivity spectra of the CCMA Nb3 samples.

### C.  Influence of wire length

Up to now we demonstrated that the dielectric response of the wire composites is significantly influenced by the local properties of microwires controlled by post-annealing techniques and doping. We also expect that the mesostructure of the wire-composite system is also of crucial importance. The lack of controlled experiments in such heterostructures leaves many unanswered questions, *e.g.* influence of the wire length. In Fig. 9, we compare the transmission parameter $S_{21}$ of the Nb1 composite containing different length of as-cast and CCMA microwires.

We now attempt to clarify the relationship between the transmission behavior and the wire length. For the composite containing 35 mm as-cast wires, we recall that the wire length is much larger than the sample width (10 mm). Consequently, it is likely that the wires form an entangled state or even bundle, during the curing cycle, rendering the dipole model no longer valid. This is consistent with the significant enhancement of conductivity (Fig. 9(a)).[38] However, the strong static magnetic interaction among wires[39] and their magnetoelastic energy are detrimental to wire entanglement. We believe the reason behind the results seen in Fig. 9 is that CCMA has an effect to significantly reduce the wire-wire magnetoelastic interactions in composite containing long wires. However, the influence of CCMA is not the full story as no blueshift of the resonance peaks is found in the transmission spectrum of the specimen filled with CCMA wires (Fig. 9(b)). A possible way to avoid wire entanglement is to select shorter wires. However, the complex nature of the $S_{21}$ spectrum for the composite containing as-cast 15 mm wires (Fig. 9(c)) precludes a precise and unique explanation. Imperfect interfacial bonding between the wires and epoxy matrix can be a possible cause for this trend. If now CCMA is performed the residual stresses in the as-cast wires are released leading to the restoration of the dielectric features (Fig. 9(d)) bearing a strong resemblance with those graphed in Fig. 4(a).

Based on the experimental results, a chemical treatment of the microwires after CCMA to modify the internal stresses and further improve the wire/epoxy interface bonding is a major avenue for future work. Recent experiments show that a silane chemical vapor deposition treatment of wires can create covalent Fe-O-Si bonding which significantly improves the interfacial conditions and soft magnetic properties.[40] Other Si-based chemical agents which have a structure consisting of hydrolysable and organically functional groups would also benefit to the bonding strength.[41]

## IV.  CONCLUDING REMARKS

In summary, we have quantified the dielectric properties of composites containing MET ferromagnetic microwires. CCMA emerges as a promising strategy to release internal stresses at the wire surface, as placed in evidence by the disappearance of the high frequency resonance peak of the effective permittivity spectra of wire composites. An interestingly strong blueshift of the low frequency resonance is observed by applying an external magnetic field to the sample containing 25 mm Nb wires. This shift can be explained by the competition between the enhanced low-field microwave absorption and the ferromagnetic resonance. Nb doping is revealed to yield a significant influence on the microstructure of the microwires, *e.g.* with 3% Nb doping, a high frequency permittivity peak at 5.0 GHz is obtained due to the surface modifications of the wire, *i.e.* appearance of a nanocrystalline phase. This Nb-induced peak is independent of CCMA in contrast to the high frequency peak identified in the Nb1 sample. At the mesoscale, the wire length is an important parameter to control the microwave permittivity. In samples containing as-cast short wires (15 mm), a complex permittivity profile is observed due to imperfect wire-epoxy bonding. In the long wires (35 mm) counterpart samples, the wires are inclined to entangle in the epoxy matrix and lead to smaller field tunability compared to that of samples with 25 mm wires. For these two kinds of sample CCMA significantly improves the permittivity results.

These findings demonstrate that the microwave properties of polymer composites containing MET wires can be controlled by the microstructure of the embedded microwires and their mesostructure. This represents a novel route towards direct, functionalizable magnetic field control of permittivity at the micro- and mesoscale, underscoring the promise of emergent electromagnetic properties in complex heterostructures.

## Acknowledgements

Y. Luo would like to acknowledge the financial support from University of Bristol Postgraduate Scholarship and China Scholarship Council. F. X. Qin is supported under the "100 Youth Talents Program" of Zhejiang University. Lab-STICC is UMR CNRS 6285.

**Figure captions**

Fig. 1 (color online) SEM image of side view of as-cast $Co_{69.25}Fe_{4.25}B_{12.5}Si_{13}Nb_1$ microwire. Each inset shows the technical details of CCMA, the cross sectional area and fracture section image of the microwires, respectively.

Fig. 2 (color online) (a) Spectra of the real part of the effective permittivity of Nb1 specimen containing 25 mm for as-cast microwires as a function of applied magnetic field; (b) As in (a) for CCMA microwires; (c) Spectra of the imaginary part of effective $\varepsilon$ of Nb1 specimen containing 25 mm for as-cast microwires as a function of applied magnetic field; (d) As in (c) for CCMA microwires.

Fig. 3 (color online) HRTEM images of as-cast and CCMA Nb1 microwires. The insets represent the fast Fourier transform (FFT) patterns of selected areas of microwires.

Fig. 4 (color online) (a) Transmission, $S_{21}$, and (c) reflection, $S_{11}$, coefficients of Nb1 composites containing 25 mm CCMA wires and their field dependence of peak position of (b) $S_{21}$ and (d) $S_{11}$, respectively.

Fig. 5 (color online) Frequency dependence of the (a) real and (b) imaginary parts of the effective permittivity of sample Nb0 containing 25 mm CCMA microwires.

Fig. 6 (color online) Frequency plots of the real part of the effective permittivity of sample Nb3 containing 25 mm (a) as-cast and (b) CCMA wires, and imaginary part of the effective permittivity of sample Nb3 containing 25 mm (c) as-cast and (d) CCMA wires.

Fig. 7 (color online) HRTEM images of as-cast and CCMA Nb3 microwires. The insets represent the FFT patterns of selected areas of microwires.

Fig. 8 (color online) Schematic of the polymer sample containing CCMA microwires with nano-clusters on the surface used to explain its microwave behavior.

Fig. 9 (color online) Transmission coefficients of Nb1 sample containing 35 mm (a) as-cast and (b) CCMA wires and 15 mm (c) as-cast and (d) CCMA wires.

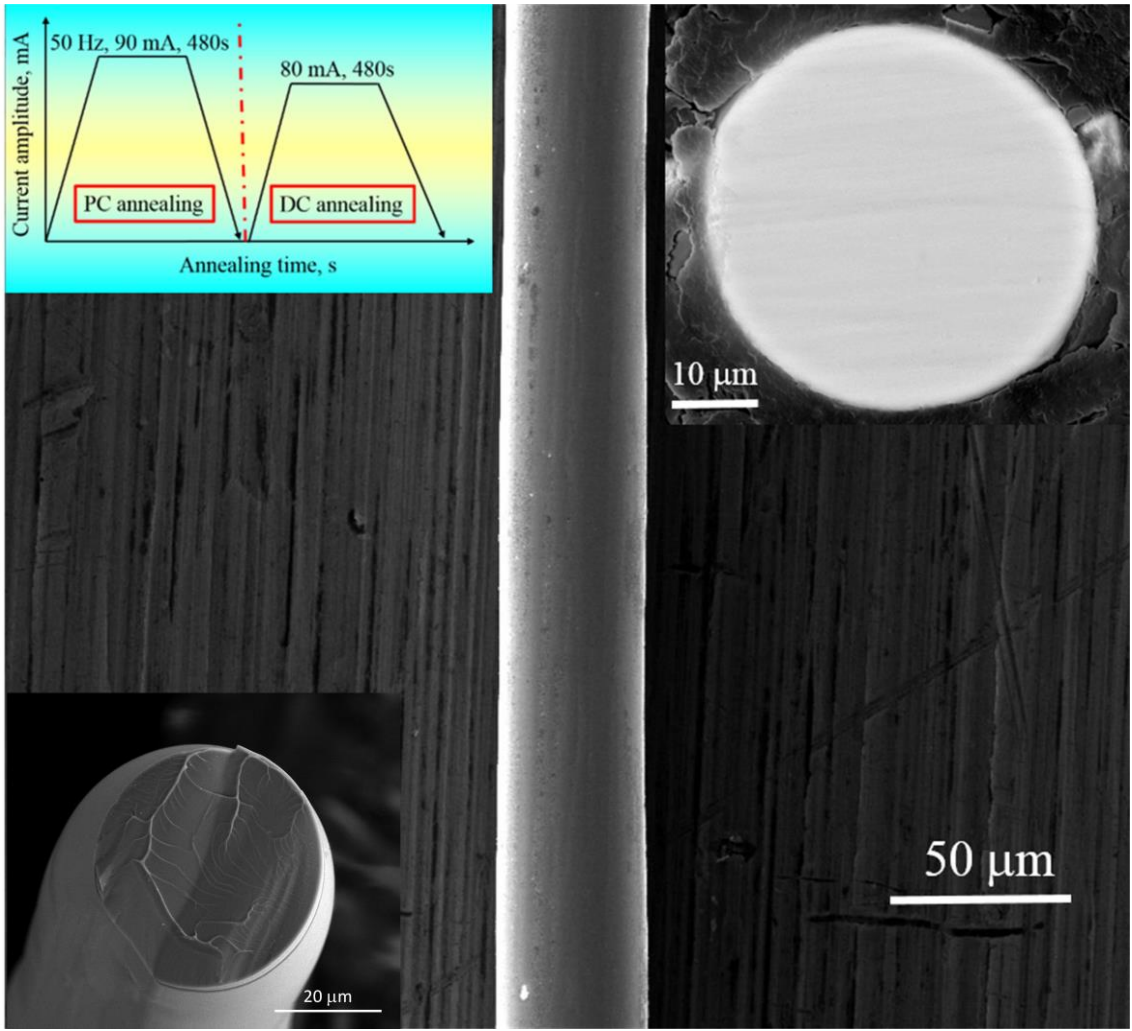

FIG. 1: Lu *et al.*

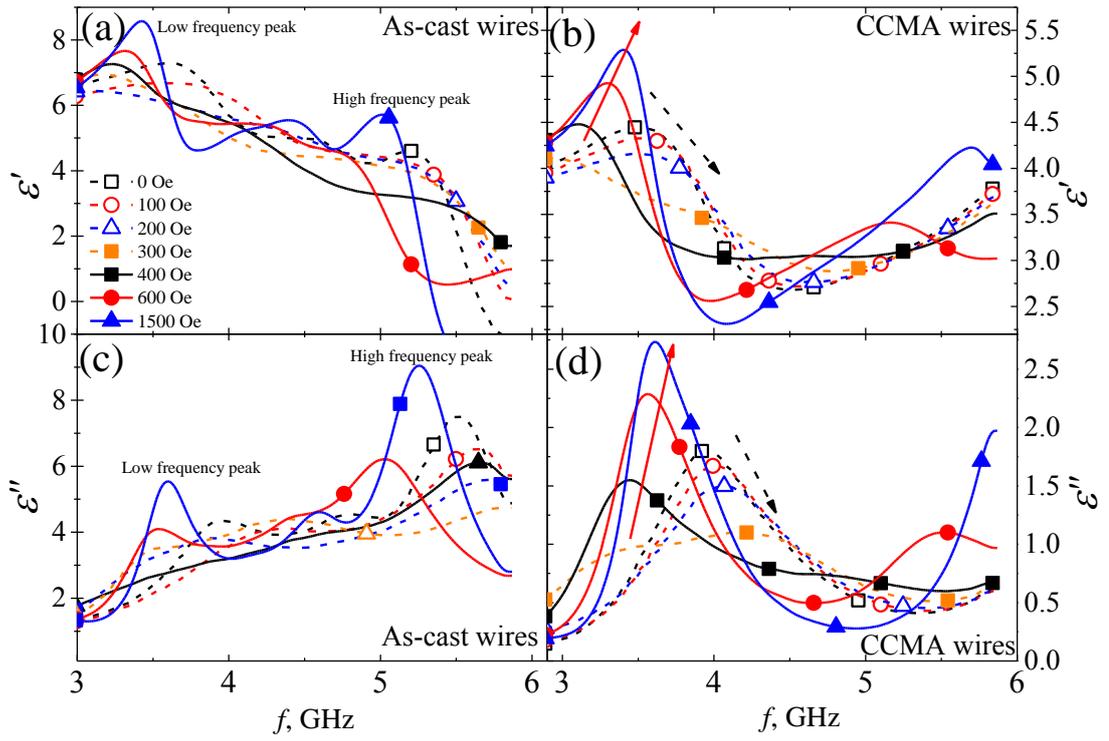

FIG. 2: Luo *et al.*

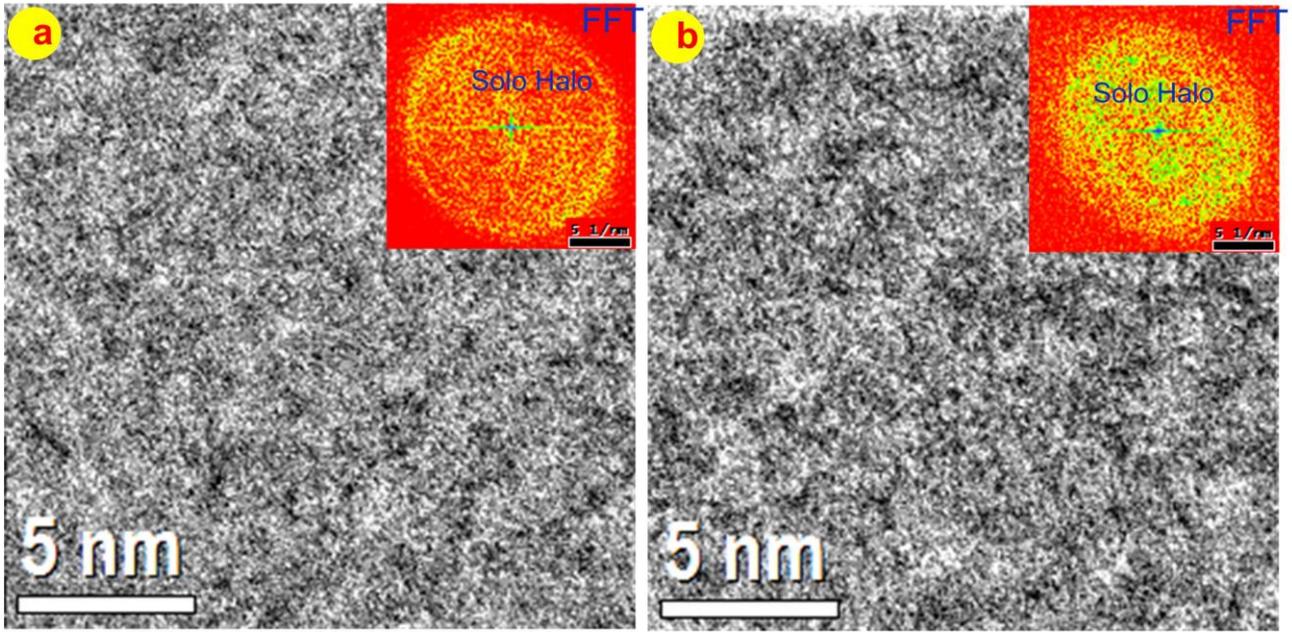

FIG. 3: Luo et al.

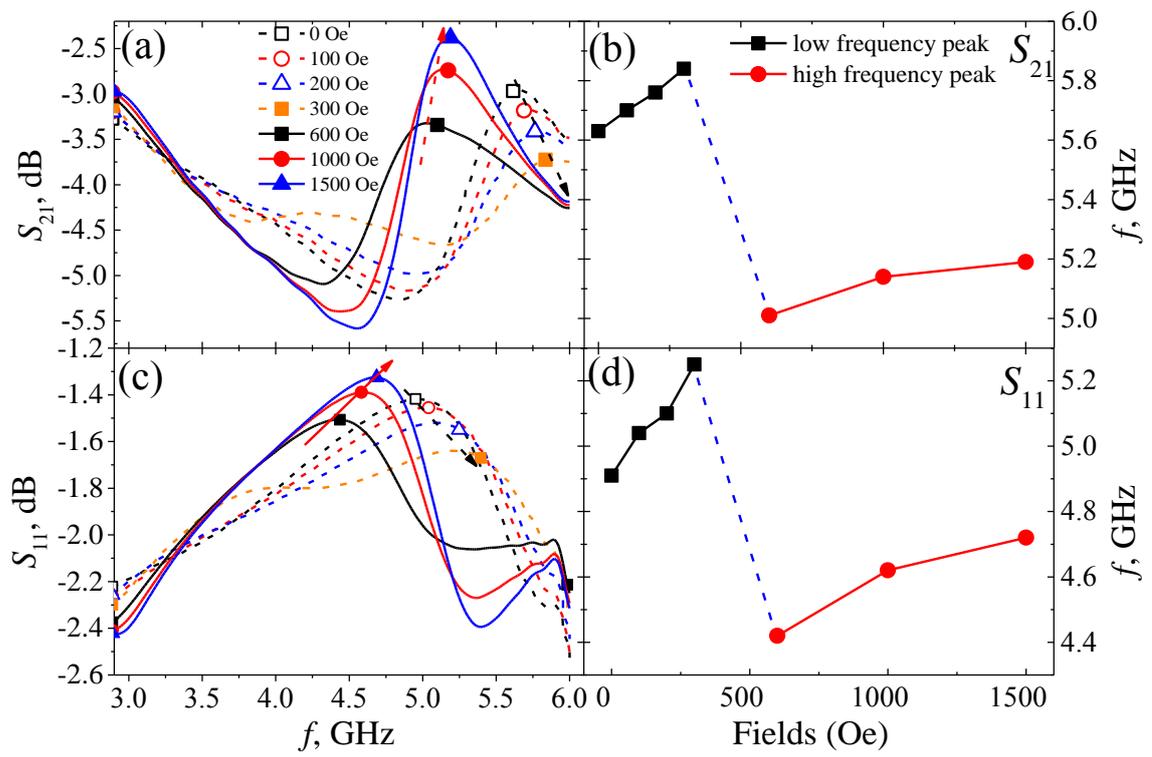

FIG. 4: Luo *et al.*

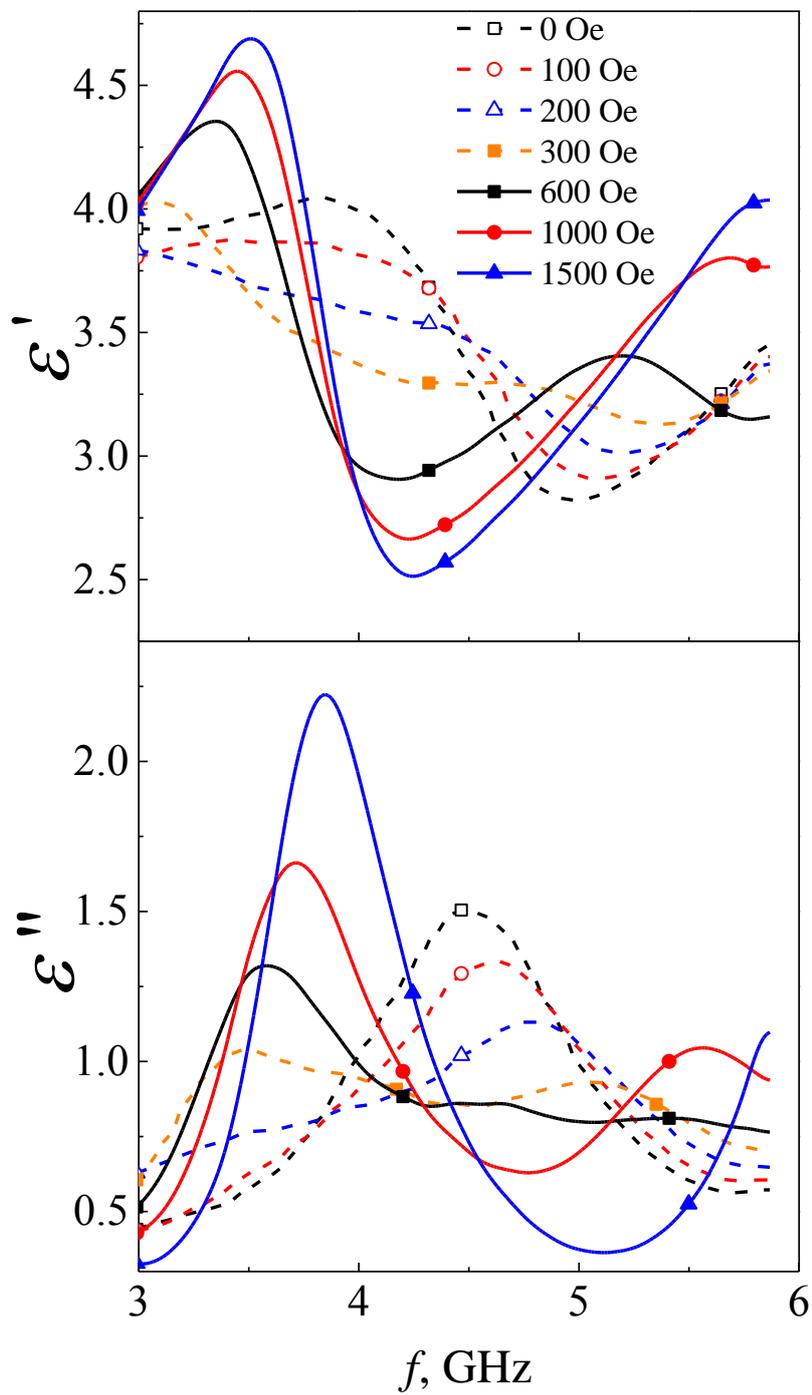

FIG. 5: Luo *et al*.

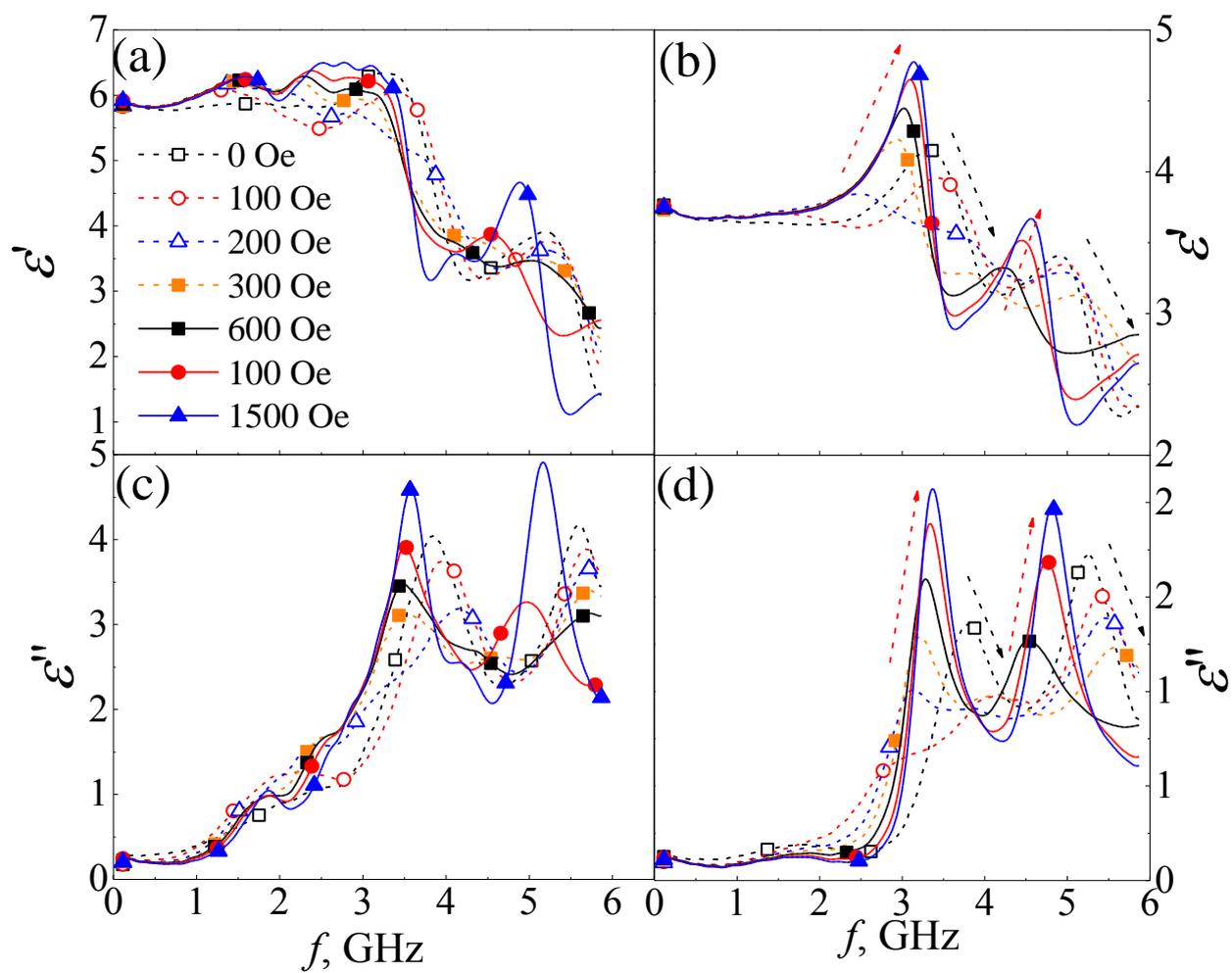

FIG. 6: Luo *et al.*

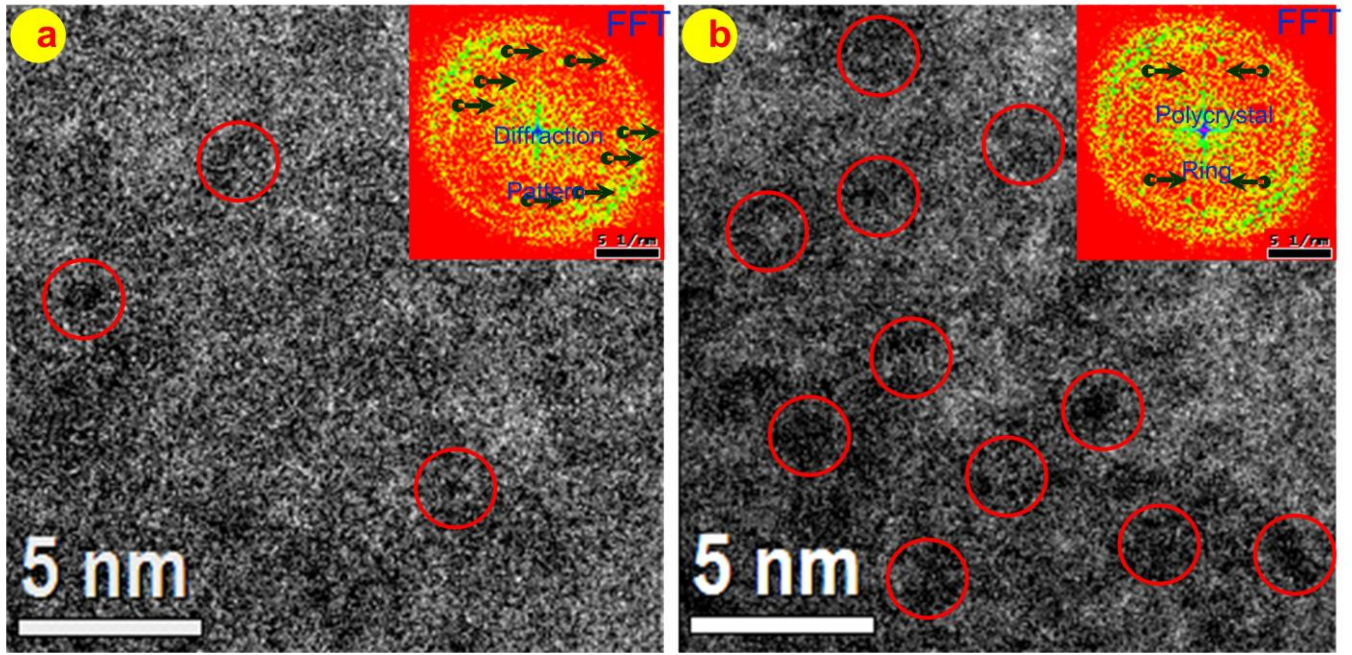

FIG. 7: Luo *et al.*

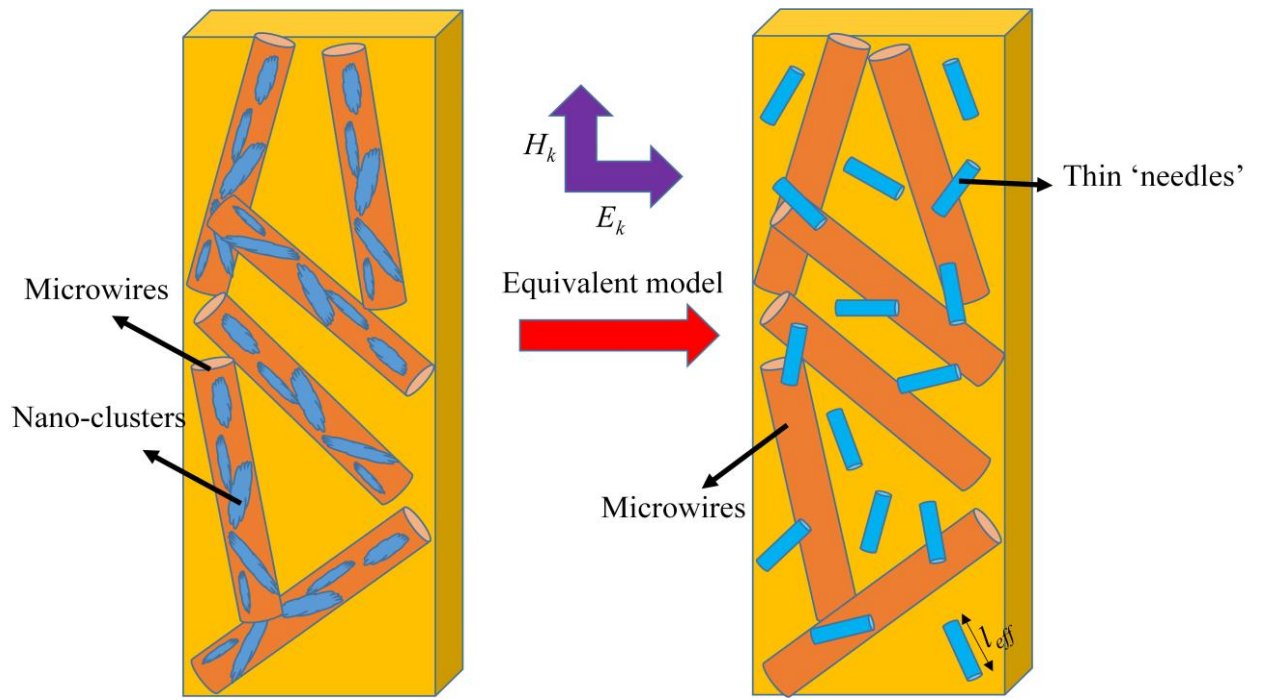

FIG. 8: Luo *et al*.

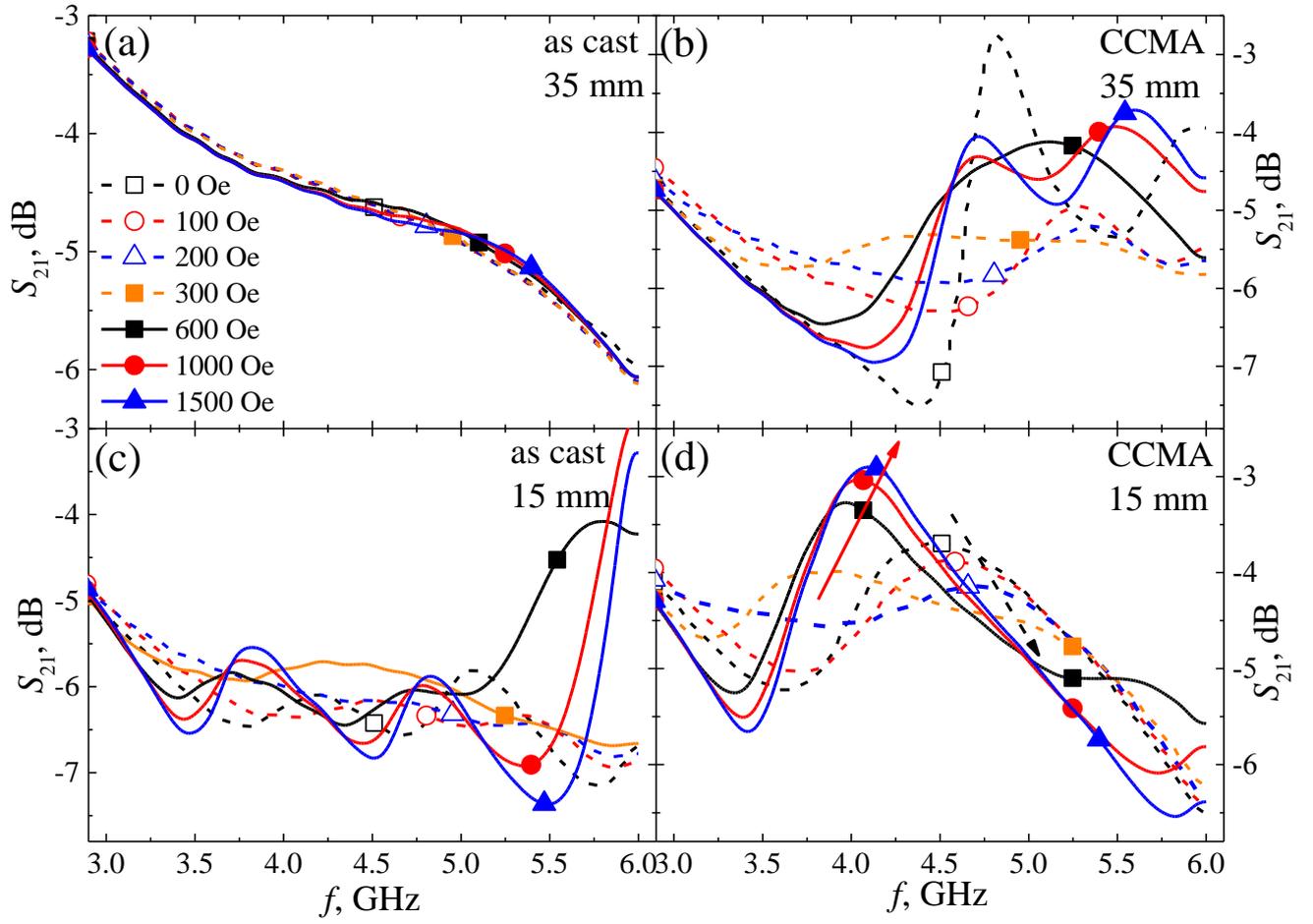

FIG. 9: Luo *et al*.